# Upper limit to magnetism in LaAlO$_3$/SrTiO$_3$ heterostructures

M. R. Fitzsimmons,[1] N. Hengartner,[1] S. Singh,[1,2] M. Zhernenkov,[1] F. Y. Bruno[3], J. Santamaria,[3] A. Brinkman,[4] M. Huijben,[4] H. Molegraaf,[4] J. de la Venta,[5] and Ivan K. Schuller[5]

[1]Los Alamos National Laboratory, Los Alamos NM 87545 USA

[2]Solid State Physics Division, Bhabha Atomic Research center, Mumbai- 400085, India

[3]GFMC. Dpto. Fisica Aplicada III, Universidad Complutense de Madrid, 28040 Madrid. Spain

[4]MESA+ Institute for Nanotechnology, University of Twente, Enschede, The Netherlands

[5]Department of Physics and Center for Advanced Nanoscience, University of California San Diego, La Jolla, CA 92093 USA

**Abstract**—Using polarized neutron reflectometry (PNR) we measured the neutron spin dependent reflectivity from four LaAlO$_3$/SrTiO$_3$ superlattices. This experiment implies that the upper limit for the magnetization induced by an 11 T magnetic field at 1.7 K is 2 emu/cm$^3$. SQUID magnetometry of the superlattices sporadically finds an enhanced moment, possibly due to experimental artifacts. These observations set important restrictions on theories which imply a strongly enhanced magnetism at the interface between LaAlO$_3$ and SrTiO$_3$.

In 2004 Ohtomo and Hwang[1] reported that the interface between $LaAlO_3$ and $SrTiO_3$ films was conducting. Since 2004 there have been numerous reports of metallic conduction,[1,2] superconductivity,[3,4] magnetism,[5] and coexistence of superconductivity and ferromagnetism[6] that have been attributed to $LaAlO_3$/ $SrTiO_3$ interfaces. On the other hand, superconductivity of oxygen deficient $SrTiO_3$ has been established for decades.[7] The conductivity in the $LaAlO_3$/ $SrTiO_3$ system may arise from intrinsic effects such as the polar catastrophe,[8,9] or extrinsic defects such as oxygen vacancies[10,11,12,13] or cation diffusion,[14,15,16] and structural distortion/orbital reconstruction.[17]

The presence of magnetic local moments at the $LaAlO_3$/ $SrTiO_3$ interface was inferred from the magnetoresistance of a (high O partial pressure grown) $LaAlO_3$/ $SrTiO_3$ interface at low temperature in high magnetic fields.[5,6,18] The location and magnitude of magnetic moments in $LaAlO_3$/ $SrTiO_3$ heterostructures remains unknown. Ferromagnetism (at room temperature), paramagnetism and diamagnetism (below 60 K) were claimed for 10 unit cells of $LaAlO_3$ on $SrTiO_3$ in a moderate field.[19] Recent torque magnetometry measurements report moments as large as $4 \times 10^{-10}$ $Am^2$ for 5 unit cells of $LaAlO_3$ on $SrTiO_3$.[20] When attributed to the entire $LaAlO_3$ film, the saturation magnetization claimed from these measurements range between 10 and 30 $emu/cm^3$—a readily detectable signal with neutron scattering.[21] If attributed to just one unit at the $LaAlO_3$ on $SrTiO_3$ interface,[20] the magnetization is noticeably larger. The large moments attributed to interfaces rely on measurement techniques that are unable to distinguish between interfacial and bulk magnetism. On the other hand, a number of experimental artifacts may produce spurious results due to extrinsic effects.[22,23] Because of this, it is imperative to perform magnetic measurements which are impervious to many experimental artifacts of which bulk measurements suffer, and to use techniques that are intrinsically sensitive to interface magnetism.

We have performed extensive polarized neutron reflectometry (PNR) measurements of LaAlO$_3$/ SrTiO$_3$ superlattices. In spite of the fact that bulk measurements of our samples (and those of others[19,20]) imply a magnetization as high as 60 emu/cm$^3$ if attributed to the superlattice, PNR unequivocally establishes an upper limit of 2 emu/cm$^3$ on their magnetization. It is important to note that the present measurements are done at fields as high as 11T and temperatures as low as 1.7K. The PNR results have serious implications for theories developed to explain magnetism in oxide superlattices.[24,25,26]

LaAlO$_3$/ SrTiO$_3$ superlattice samples were grown independently at laboratories located in Twente and Madrid. The Twente samples were grown on TiO$_2$ terminated SrTiO$_3$ single crystal substrates. [(LaAlO$_3$)$_8$/(SrTiO$_3$)$_{24}$]$_{30}$ (Twente1) and [(LaAlO$_3$)$_4$/(SrTiO$_3$)$_{12}$]$_{30}$ (Twente2) superlattices were fabricated by pulsed laser deposition with reflection high-energy electron diffraction (RHEED) control of the growth process. Single-crystal LaAlO$_3$ and SrTiO$_3$ targets were ablated at a laser fluence of 1.3 J/cm$^2$ and a repetition rate of 1 Hz. During growth, the substrate was held at 850 °C in an oxygen environment at 2x10$^{-3}$ mbar for LaAlO$_3$ and SrTiO$_3$. RHEED intensity oscillations were observed during growth of each individual layer indicating control on the unit cell scale due to the layer-by-layer growth mode. After growth, the superlattices were cooled to room temperature in 2x10$^{-3}$ mbar of oxygen at a rate of 10 °C/min. Atomic force microscopy (AFM) showed smooth terraces separated by unit cell high steps similar to the surface of the initial TiO$_2$-terminated SrTiO$_3$ (100) substrate.

[(LaAlO$_3$)$_8$/(SrTiO$_3$)$_{24}$]$_{30}$ (Madrid2) and [(LaAlO$_3$)$_4$/(SrTiO$_3$)$_{12}$]$_{15}$ (Madrid1) were grown on TiO$_2$ terminated (001) STO substrates using high-pressure (2.9 mbar) pure oxygen sputtering at a substrate temperature of 750 °C. After growth the samples were cooled to 600 °C. At this temperature the chamber was filled with 900 mbar O$_2$, and the samples were annealed for 5 minutes at 550 °C before completing the cooling to room temperature at a 20 °C/min rate.

The physical structure is determined from the wavevector transfer (Q)-dependence of the X-ray reflectivity which is sensitive to the uniformity of layer thicknesses and interface roughness averaged over lateral dimensions of several microns. The broadening of the superlattice reflections implies that the layer thicknesses varied across the sample (along its surface normal) by more than half a unit cell. Furthermore, the positions of the superlattice reflections for the Madrid1 and Twente2 samples were shifted slightly towards larger Q compared to the simulation implying the LaAlO$_3$ layer thicknesses for these samples may be (3.5 ± 0.5) unit cells thick rather than 4 unit cells as intended. On the other hand, the LaAlO$_3$ thickness of Samples Twente1 and Madrid2 were closer to the intended thickness of 8 unit cells.

To test whether our LaAlO$_3$/SrTiO$_3$ heterostructures are magnetic, we measured the depth dependence of magnetization for several LaAlO$_3$/ SrTiO$_3$ superlattices at different fields and temperatures with polarized neutron reflectometry (PNR).[27,28,29] In PNR the intensity of specularly reflected neutrons is compared to the intensity of the incident beam as a function of Q and neutron beam polarization. The specular reflectivity, R, is determined by the neutron scattering length density depth profile, $\rho(z)$, averaged over the lateral dimensions of the

sample. ρ(z) consists of nuclear and magnetic scattering length densities such that

, where C = 2.853x10$^{-9}$Å$^{-2}$cm$^3$/emu and M(z) is the depth profile of the magnetization (in emu/cm$^3$) parallel to the applied field.[29] The +(-) sign denotes neutron beam polarization parallel (opposite) to the applied field. Thus, by measuring R$^+$(Q) and R$^-$(Q), and        can be obtained separately. The difference between R$^+$(Q) and R$^-$(Q) or the difference divided by the sum, the "spin asymmetry", is very sensitive to small M values. A major advantage of PNR is the ability to distinguish magnetism at an interface or a thin film, from the substrate. Measurements of the spin-dependent superlattice reflection assure that variations of M having the period of the superlattice and not spurious contamination from other sources are measured. Thus, we should stress that the important concerns raised regarding bulk magnetometry of materials with nanometer dimensions and phantom magnetism[23] are not germane to PNR.

Detection of small M can be problematical even for PNR. First, the difference between R$^+$ and R$^-$ may be so small that the difference is not statistically significant. This difference can be substantially enhanced using a superlattice, where many interfaces contribute to the superlattice reflection. Additionally, a neutron scattering instrument may bias one spin state over the other. For example, to reverse the neutron beam polarization a neutron spin flipper is turned off or on, thus treating the two polarizations differently which may induce a systematic error on R$^+$ or R$^-$. Previously, we have shown the instrumental bias for the Asterix reflectometer/diffractometer is less than one part in 1000.[30]

In order to further suppress instrumental bias, we developed a new measurement protocol for this experiment. Two samples measured at the same time were compared—a LaAlO$_3$/ SrTiO$_3$ superlattice, and a control sample (in this case a MgO single crystal substrate). The samples were mounted on a special holder with approximately 1° difference between their surface normals [Figures 1(a) and 1(b)]. Thus, the specularly reflected beams corresponding to the superlattice reflection from the LaAlO$_3$/ SrTiO$_3$ and the region of total reflection from the control sample appear in different locations of a position sensitive neutron detector [Figure 1(c)]. The neutron intensity was measured as a function of wavelength $\lambda$ for fixed scattering angle $2\theta$ from which we obtain ———. The incident beam intensity for each spin state was determined using a portion of the spectrum of the neutron beam reflected by the control sample. This protocol normalizes out instrumental artifacts that might produce a false spin asymmetry, since the sample and control are measured simultaneously.

Samples and control were cooled in an 11 T field from room temperature to 1.7 K. The spin dependent reflectivities were simultaneously recorded as functions of field and temperature using the Asterix spectrometer at LANSCE. The data were corrected for variation of the neutron spectrum and wavelength dependent variation of the efficiencies of the neutron polarizer and spin flipper.[31]

Figure 2(a) shows the reflectivity of Sample Twente1 near its critical edge (where the sample's reflectivity is unity) and over a range of Q which includes the Bragg reflection for Sample Madrid1 (inset). The Q-dependent spin asymmetry near the critical edge near $Q_c$ = 0.0133 Å$^{-1}$ for SrTiO$_3$, which is sensitive to the net magnetization, is shown for Sample Twente1 in Figure 2(b). These data were analyzed using the method of Parratt[32] assuming a uniformly distributed magnetization depth profile. A magnetization (-0.12 ± 1.2) emu/cm$^3$ was obtained implying a statistically insignificant spin asymmetry at the critical edge of the LaAlO$_3$/ SrTiO$_3$. The influence on the spin asymmetry of magnetization equal to -0.12, 1, and 10 emu/cm$^3$ uniformly distributed throughout the superlattice are shown by the red, dashed and blue curves in Figure 2(b), respectively.

To quantify influence of field and temperature on the spin dependence of the Bragg reflections from the LaAlO$_3$/ SrTiO$_3$ [an example is shown in the inset of Figure 2(a)], we define a spin asymmetry ratio (SAR). This ratio is equal to the difference between the *integrated intensities* of the spin dependent superlattice reflections divided by their sum.[33] Figure 3 shows the SAR for Sample Twente1 as a function of field for two temperatures averaged from two separate experiments. Figure 4 shows the SAR for four samples measured at 11 T and 1.7 K.

The spin asymmetry of the superlattice reflection is sensitive to changes of magnetization having the period of the superlattice. It is possible for such changes to be too small to produce spin asymmetry at the critical edge. For example, the magnetization may change sign as a function of depth, or it may be confined to an interface, i.e., to a small fraction of the superlattice.

In Figure 3 we define the slope from the best fits to the data weighted according to the 1-sigma errors which gives a 1-sigma error on the slope. The slope of the 1.7 K data (closed symbols, Figure 3), (-0.0006 ± 0.0002) T$^{-1}$ is statistically different than zero and suggests that *the*

*SAR becomes more negative as the field increases*. On the other hand, the slope for the 80 K data (open symbols, Figure 3), (-0.0003±0.0003) $T^{-1}$ is zero within statistical error. These results imply that the influence of field on *the SAR diminishes with increasing temperature*.

In order to determine an upper limit on the change of magnetization across the $LaAlO_3$/ $SrTiO_3$ superlattices, we established a relation between the SAR and magnetization in absolute units. The SAR for our samples depends linearly on the change of magnetization across the $LaAlO_3$/ $SrTiO_3$ interface. This relationship can be used to calibrate the right hand axes of Figures 3 and 4. If the magnetizations are parallel to the applied field a negative SAR implies that the magnetization in $SrTiO_3$ is greater than that of $LaAlO_3$. For the Sample Twente1 at 11 T and 1.7 K, the SAR is about -0.005 which implies that the $SrTiO_3$ magnetization is ~2 emu/cm$^3$, if the magnetization of $LaAlO_3$ is assumed to be zero. As a consequence, the thickness-weighted average magnetizations of the $SrTiO_3$ and $LaAlO_3$ layers, are below the upper limit set by the absence of spin asymmetry at the critical edge of the sample.

Alternatively, if the magnetization is opposite to the applied field, i.e., diamagnetic, a negative SAR implies that the magnetization resides in the $LaAlO_3$ layer. Our measurements cannot distinguish between induced paramagnetism in $SrTiO_3$, or induced diamagnetism in $LaAlO_3$.

Overall the SAR of the superlattice reflection in combination with measurements near the critical edge imply that: the magnetization in the $LaAlO_3$ layer is ~2 emu/cm$^3$ less than that of the $SrTiO_3$ layer (or some part thereof), the magnetization of one of these layers is near zero, and the variation of magnetization with depth has the period of the superlattice. If attributed to

Ti, the upper limit for the magnetic moment of 2 emu/cm$^3$ would correspond to 0.7% of Ti$^{3+}$ per unit cell. This value is of the same order as the few percent of Ti 3d$_{xy}$ electrons reported by x-ray spectroscopy.[17,34] The magnetic scattering centers inferred by Brinkman et al.[5] may originate from interface Ti$^{3+}$ ions.

The magnetization in LaAlO$_3$/SrTiO$_3$ in our samples is unquestionably small and is not consistent with the large magnetic moments that SQUID (ours or those of Ref. [19]) and torque magnetometry[20] attribute to the interface. This discrepancy can either be explained by different growth conditions (magnetotransport effects are most easily observed in samples grown under high O pressure [5,19] such as used by us), by different interface reconstruction at multilayers compared to bilayers, by bulk contributions to the magnetism, or by use of bulk magnetometry techniques that are prone to artifacts.[22,23]

In conclusion, we established an upper limit of 2 emu/cm$^3$ for the change of magnetization across the LaAlO$_3$/SrTiO$_3$ interfaces in superlattices at 11 T and 1.7 K. The upper limit was obtained from measurements of differences between the integrated superlattice Bragg intensities for neutron beam polarization parallel and opposite to the field applied. No significant spin difference was measured near the critical edge of the LaAlO$_3$/SrTiO$_3$ superlattice. Thus, the magnetization *averaged over the entire superlattice* is likely to be less than the upper limit of ~1 emu/cm$^3$ inferred from measurement of the MgO control, and certainly less for fields less than 11 T or temperatures greater than 1.7 K. These results are inconsistent with magnetization of the order of tens of emu/cm$^3$ for free electron spin densities of $10^{21}$ cm$^{-3}$ aligned by an 11-T field. Some theories of conductivity predict electron spin densities of this magnitude at or near the interface between LaAlO$_3$ and SrTiO$_3$.

**Acknowledgements**

This work was supported by the Office of Basic Energy Science, U.S. Department of Energy, BES-DMS funded by the Department of Energy's Office of Basic Energy Science. Los Alamos National Laboratory is operated by Los Alamos National Security LLC under DOE Contract DE-AC52-06NA25396.  Work at UCM is supported by Consolider Ingenio CSD2009-00013 (IMAGINE), CAM S2009-MAT 1756 (PHAMA) and work at Twente is supported by the Foundation for Fundamental Research on Matter (FOM).

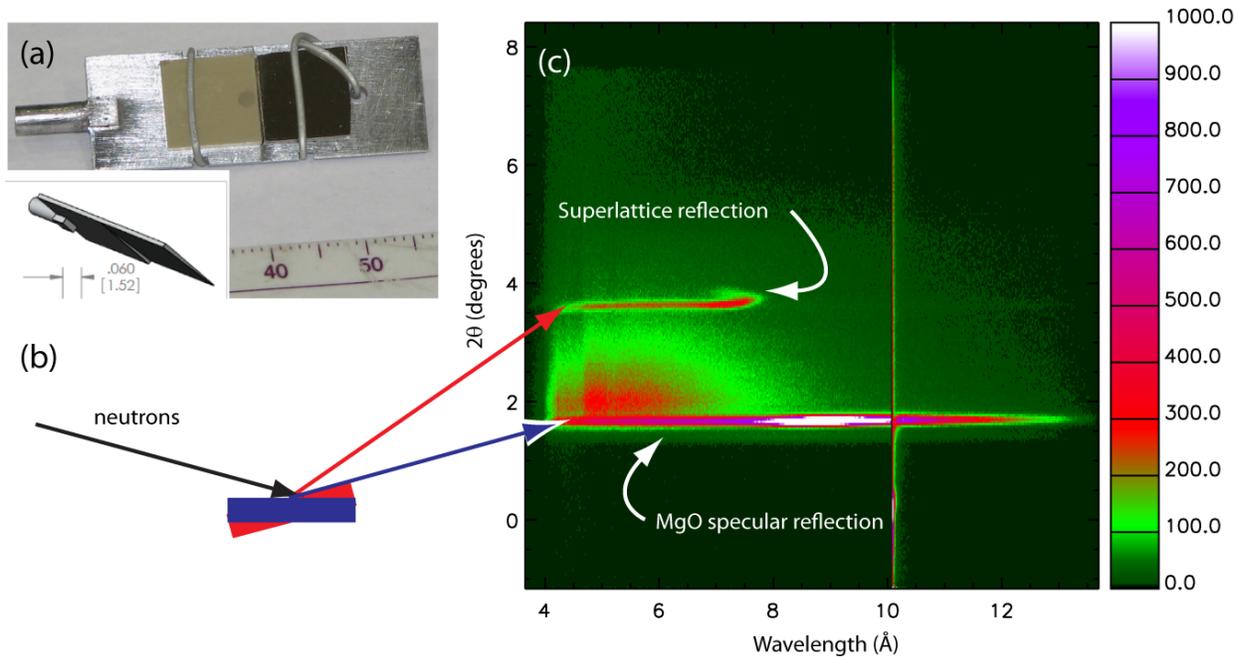

**Figure 1** (a) Device used to simultaneously hold the control and LaAlO$_3$/ SrTiO$_3$ samples during the PNR experiment. (b) Schematic showing reflection of the neutron beam by the control (blue) and LaAlO$_3$/ SrTiO$_3$ sample (red). (c) Neutron intensity image for one beam polarization.

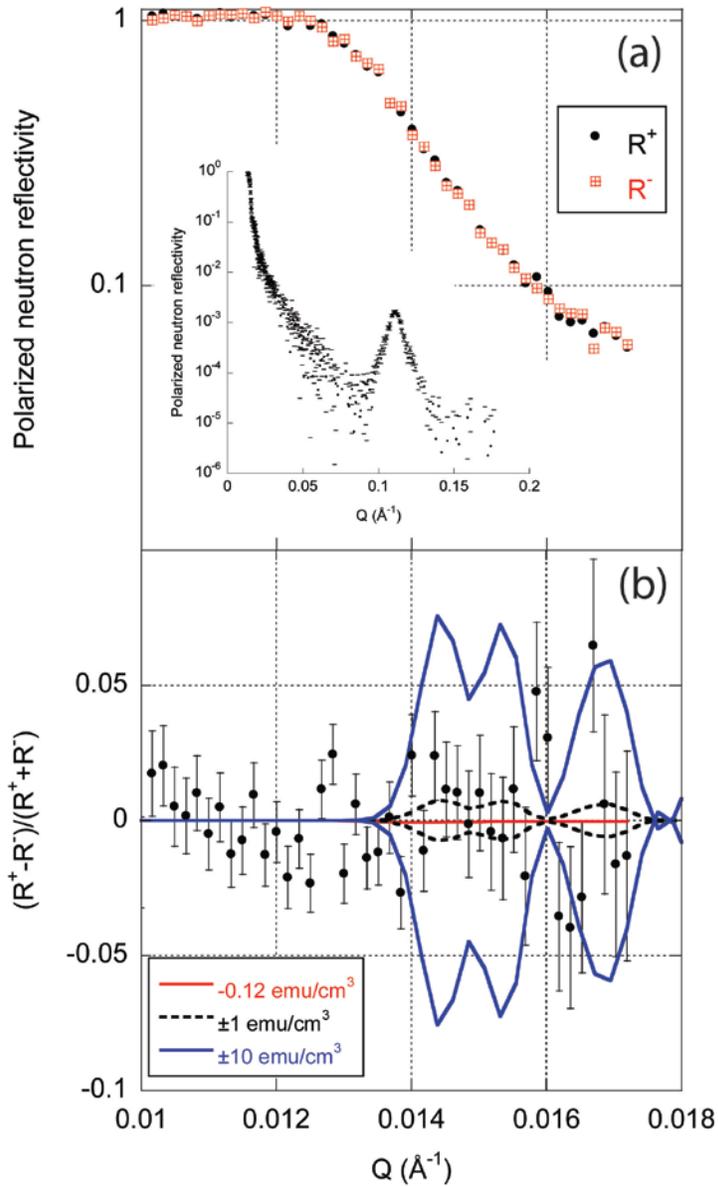

**Figure 2** (a) Polarized neutron reflectivity from Twente1 near the critical edge and (inset) $R^+$ from Madrid1 over a broader range of Q showing the superlattice reflection at $Q = 0.12\text{Å}^{-1}$. (b) The neutron spin asymmetry of Sample Twente1 near the critical edge. (1-sigma errors) Curves correspond to the influence on the spin asymmetry of magnetization uniformly distributed in the superlattice.

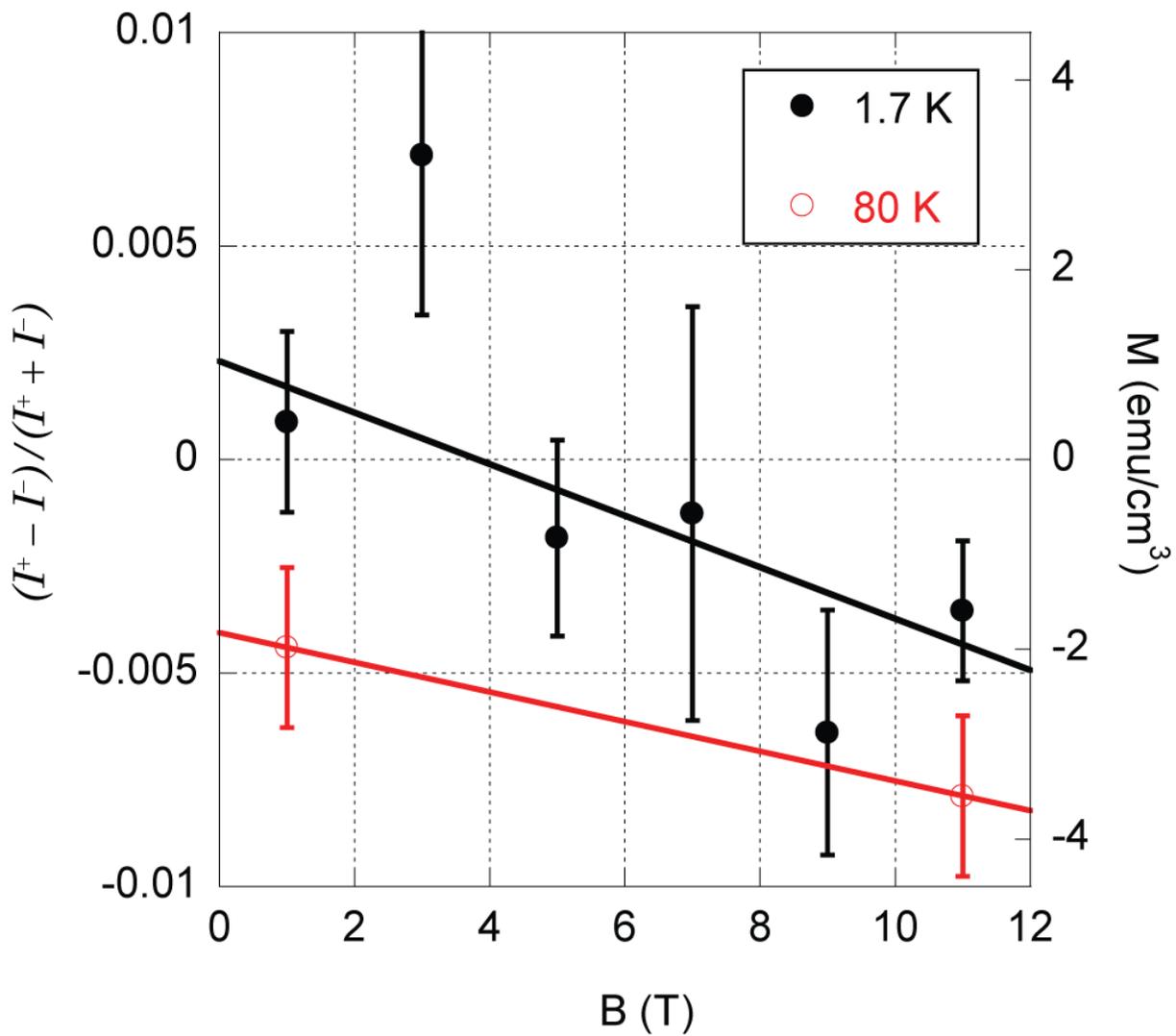

**Figure 3** Ratio of the difference over the sum of the integrated intensities, SAR, of the spin dependent superlattice reflections for Twente1 as functions of field and temperature. (1-sigma errors)

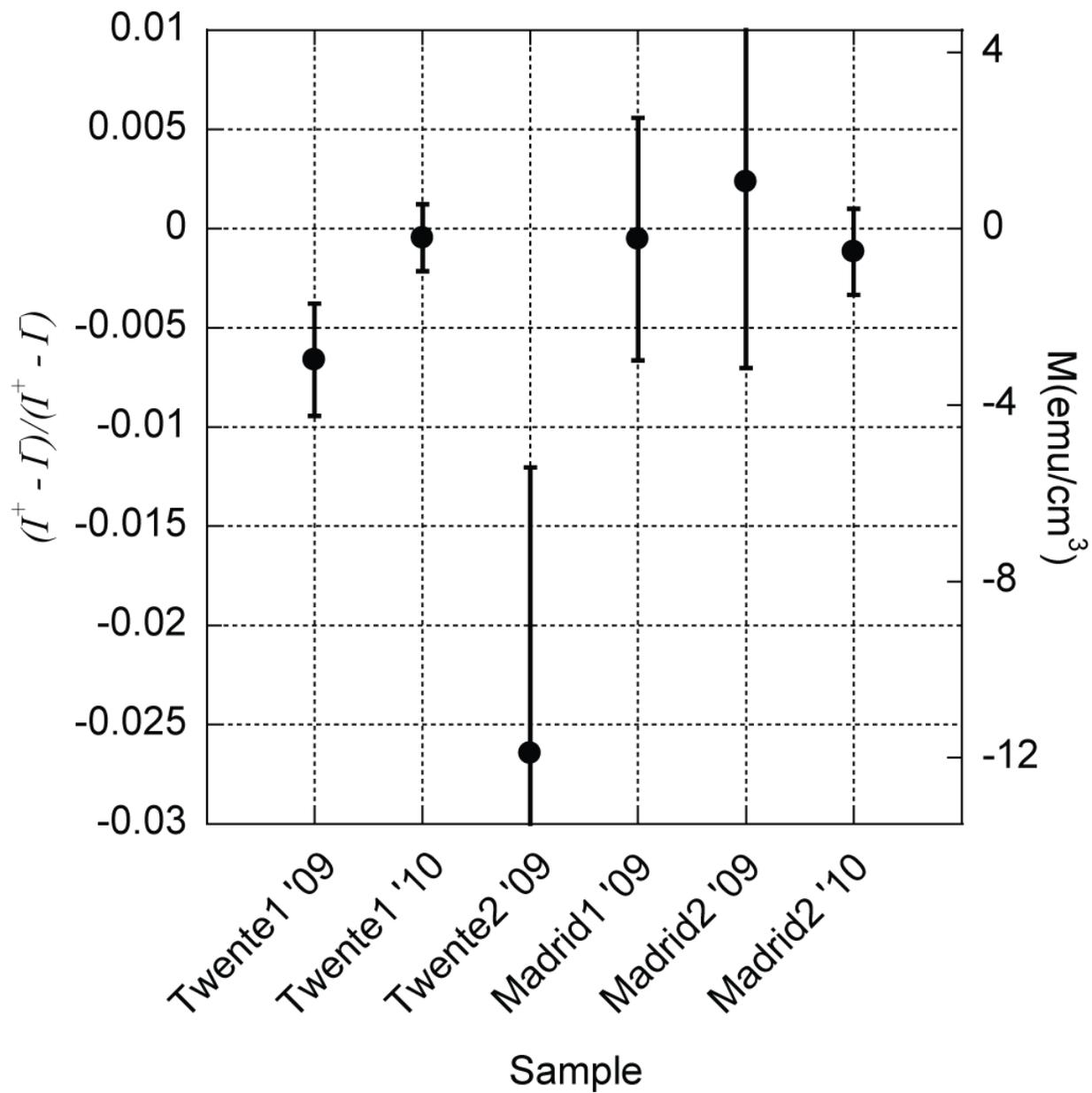

**Figure 4** Ratio of the difference over the sum of the integrated intensities, SAR, of the spin dependent superlattice reflections for different samples at 11 T and 1.7 K. (1-sigma errors)